\documentclass[a4paper]{jpconf}
\usepackage{graphicx}
\usepackage{hyperref}
\usepackage{amsmath,amssymb}
\usepackage{dsfont}
\newcommand{\sh}[1]{#1\hskip-7pt \diagup}
\begin{document}
\title{Unsterile-Active Neutrino Mixing}

\author{Jimmy A. Hutasoit\footnote{This talk is based on a work with D. Boyanovsky and R. Holman \cite{Boyanovsky:2009ke}.}}

\address{Department of Physics, Carnegie Mellon University, Pittsburgh, PA 15213, USA.}

\ead{jhutasoi@andrew.cmu.edu}

\begin{abstract}
We consider a sterile neutrino to be an unparticle, namely an ``unsterile'' neutrino, and study its mixing with a canonical active neutrino via a see-saw mass matrix. There is no unitary transformation that diagonalizes the mixed propagator and a field redefinition is required. The unsterile-like propagating mode features a resonance for anomalous dimension between 0 and 1/3, but the complex pole disappears when the anomalous dimension is larger than 1/3. The active-like propagating mode is described by a stable pole, but inherits a non-vanishing spectral density above the unparticle threshold. We also find that the radiative decay width of the unsterile neutrino into the active neutrino (and a photon) via charged current loops is suppressed, and this suppression weakens the bound from the X-ray or soft gamma-ray background when one considers the unsterile neutrino to be a dark matter candidate.
\end{abstract}

\section{Introduction}
With the advancement of string theory, in particular with the advent of the much celebrated gauge/gravity duality, conformal invariance, along with supersymmetry, enjoys a lot of attention from the theoretical physics community. However, unlike supersymmetry, the pursuit of the phenomenological implications of having a conformal invariant sector in our universe had been lacking. This is not surprising, considering that in our everyday infrared life, size does matter and particles do have definite masses. 

Recently, Georgi suggested an extension of the standard model in which particles couple to a \emph{hidden} conformal sector \cite{georgi}. At low energy, there emerges an effective interpolating field that features an anomalous scaling dimension. Georgi called this field the \emph{unparticle} field as it has a phase space that looks like those of particles, but with a non-integer number of particles.

In this talk, we will consider the possibility of a fermionic unparticle that is not charged under any gauge groups of the standard model, but that is mixed with the standard model neutrinos, or the active neutrinos, via a see-saw type mass matrix. Henceforth, we will call this fermionic unparticle the \emph{unsterile} neutrino.

\section{The Set-up}
Let us consider a simple set-up that consists of an unsterile and an active Dirac neutrino, mixed via a see-saw mass matrix. In momentum space, the Lagrangian density is given by
\begin{equation}
\mathcal{L} = \overline{\psi}_U \; \left(\sh{p} -M\right) \,F(p)\;\psi_U + \overline{\nu}_a\; \sh{p}\; \nu_a - m \; \left(\overline{\psi}_U \, \nu_a + \overline{\nu}_a \, \psi_U \right), \label{unlag}
\end{equation}
where
\begin{equation}
F(p) = \left[\frac{-p^2+M^2-i\epsilon}{\Lambda^2}\right]^{-\eta}\, ,
\end{equation}
with $0<\eta<1$. Here, $\psi_U$ and $\nu_a$ are the unsterile and the active neutrinos, respectively.

\begin{figure}[h]
\begin{center}
\includegraphics[width= 8 cm,keepaspectratio=true]{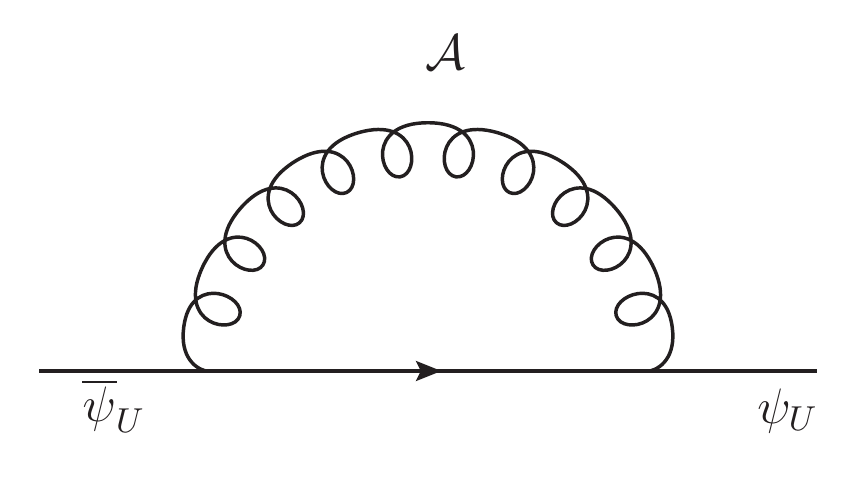}
\caption{The leading order contribution to the self-energy.}
\label{fig:selfenergy}
\end{center}
\end{figure}

The kinetic term for the unsterile can be understood in many ways, but let us explain it in the language of RG resummation \cite{bogo, infradiv,Neubert:2007kh}. Let us consider the unsterile to be interacting with a field $\mathcal{A}$ of the conformal hidden sector by an interaction of the form
\begin{equation}
\mathcal{L}_{\rm int} = g\, \overline{\psi}_U \mathcal{A} \psi_U, \label{hidden}
\end{equation}
where $g$ is a small dimensionless coupling. To lowest order in $g$, the self-energy of $\psi_U$ is given by
\begin{equation}
\Sigma (p) = -\eta\, \left(\sh{p} -M\right)\,\ln\left[\frac{-p^2+M^2-i\epsilon}{\Lambda^2}\right],
\end{equation}
where $\eta=cg^2$, with $c$ is a constant whose value depends on the nature of the conformal field $\mathcal{A}$, and $\Lambda$ is a renormalization scale. We have also subtracted the self-energy such that it vanishes at $\sh{p}=M$. The Feynman diagram for the self-energy is depicted in Fig. \ref{fig:selfenergy}.

Integrating out the conformal field $\mathcal{A}$ leads to the following effective action for $\psi_U$
\begin{eqnarray}
\mathcal{L}_{\rm eff} &=& \overline{\psi}_U\,\left(\sh{p} -M\right)\Bigg[1-\eta \ln\left(\frac{-p^2+M^2-i\epsilon}{\Lambda^2}\right)\Bigg] \psi_U \nonumber \\
&\approx& \overline{\psi}_U\,\left(\sh{p} -M\right)\Bigg[ \frac{-p^2+M^2-i\epsilon}{\Lambda^2}\Bigg]^{-\eta} \psi_U,
\end{eqnarray} 
where in going to the last line we have invoked a renormalization group resummation of the infrared threshold divergences.
 
\section{Diagonalization and the Results}
By introducing the ``flavor" doublet
\begin{equation}
\Psi =  \Big(\begin{array}{c}
                     \nu_a \\
                     \psi_U
                   \end{array}\Big), 
\end{equation}
we can rewrite the Lagrangian density in a more compact form
\begin{equation}
\mathcal{L} = \overline{\Psi}(-p)\,\Big[\sh{p} \, \mathds{F} - \mathds{M} \Big]\,\Psi(p), 
\end{equation}
where
\begin{equation}
\mathds{F} = \Bigg( \begin{array}{cc}
                                                  1 & 0 \\
                                                  0 & F(p)
                                                \end{array}
\Bigg)
\qquad {\rm and} \qquad
\mathds{M} =  \Bigg( \begin{array}{cc}
                                                  0 & m \\
                                                  m & M F(p)
                                                \end{array}
\Bigg).
\end{equation}

To gain insight into how we can diagonalize the Lagrangian density, it is instructive to rewrite it in the helicity basis
\begin{equation}
{\cal L} =  \sum_{h=\pm 1} \left( \begin{array}{cc}
                        {\Psi_R^h}^{\dagger } &  {\Psi_L^h}^{\dagger}
                      \end{array} \right) \,  \left( \begin{array}{cc}
                        (p^0 - h |\vec{p}|) \mathbb{F} &  \mathbb{M}\\
                        \mathbb{M} &  (p^0 + h |\vec{p}|) \mathbb{F}
                      \end{array} \right) \, \left( \begin{array}{c}
                        \Psi_R^h \\  \Psi_L^h
                      \end{array} \right).
\end{equation}                      
As the action is manifestly Lorentz invariant, the transformation we need to perform in order to diagonalize the action must be such that the final result is still Lorentz invariant. In particular, Lorentz invariance does not allow us to mix the $(2,1)$-representation of the Lorentz group with the $(1,2)$-representation. Therefore, to diagonalize the action, we are going to perform the transformation $\Psi_{R,L}^h \rightarrow {\cal U}_{R,L} \, \Psi_{R,L}^h$, such that all the following matrices:
\begin{equation}
 {\cal U}_{R} \,\Big[ (p^0 - h |\vec{p}|) \mathds{F}\Big] \, {\cal U}_{R}^{-1},~~~~ {\cal U}_{L} \,\Big[ (p^0 + h |\vec{p}|) \mathds{F}\Big] \, {\cal U}_{L}^{-1},~~~~
 {\cal U}_{R} \;\mathds{M} \; {\cal U}_{L}^{-1}~~~~{\rm and}~~~~{\cal U}_{L} \;\mathds{M} \; {\cal U}_{R}^{-1}  \label{umu}
\end{equation}
are all diagonal. Since $\mathds{F}$ is diagonal, and yet not proportional to the identity, the only possible unitary transformations that diagonalize the first two are
\begin{equation}
{\cal U}_{R,L} =  \left( \begin{array}{cc}
                        1 &  0\\
                       0 &  \pm \, 1
                      \end{array} \right) ~~ {\rm or} ~~ \left( \begin{array}{cc}
                        \pm \, 1 &  0\\
                       0 &  1
                      \end{array} \right).
\end{equation}
However, none of the combinations of these possibilities diagonalize the last two matrices in  \ref{umu}. Therefore, there is no unitary transformation that diagonalizes the full propagator.

Instead, we are going to diagonalize the action by first rescaling the unsterile field, such that the flavor doublet is now given by
\begin{equation}
\nu = \Bigg(\begin{array}{c}
             \nu_a \\
             \nu_U
           \end{array} \Bigg),
\end{equation} 
with $\nu_U = \sqrt{F(p)}\, \psi_U$, and the Lagrangian density becomes
\begin{equation}
\mathcal{L} = \overline{\nu}\, \Big[\sh{p} \,\mathds{I} - \widetilde{\mathds{M}}\Big]\, \nu,
\end{equation}
where
\begin{equation}
\widetilde{\mathds{M}} = \frac{1}{\sqrt{\mathds{F}}} ~ \mathds{M} ~ \frac{1}{\sqrt{\mathds{F}}} = \Bigg( \begin{array}{cc}
         0 & \frac{m}{\sqrt{F(p)}} \\
         \frac{m}{\sqrt{F(p)}} & M
       \end{array}
 \Bigg),
\end{equation}
and $\mathds{I}$ is the identity in flavor space.

It is now pretty straightforward to diagonalize the action. The propagating or ``mass" eigenstates are given by
\begin{equation}
\Bigg( \begin{array}{c}
            \nu_a\\
             \nu_U
           \end{array}
\Bigg) =   U(p)   \Bigg( \begin{array}{c}
             \nu_1 \\
             \nu_2
           \end{array}
\Bigg),
\end{equation}
with
\begin{equation}
U(p) = \frac{1}{\sqrt{2}} \, \left(
                      \begin{array}{cc}
                         \Big[1+\widetilde{\mathcal{C}}(p)\Big]^\frac{1}{2}  &  \Big[1-\widetilde{\mathcal{C}}(p)\Big]^\frac{1}{2}  \\
                        - \Big[1-\widetilde{\mathcal{C}}(p)\Big]^\frac{1}{2}  &  \Big[1+\widetilde{\mathcal{C}}(p)\Big]^\frac{1}{2}  \\
                      \end{array}
                    \right)
\qquad {\rm and} \qquad
\widetilde{\mathcal{C}}(p) =  \Bigg[1+ \frac{4 m^2}{M^2 F(p)} \Bigg]^{-\frac{1}{2}}.
\end{equation}
The diagonalized ``mass" matrix is then given by
\begin{equation}
\widetilde{\mathbb{M}}_d = U^{-1}(p) ~\widetilde{\mathbb{M}}~U(p) = \Bigg(\begin{array}{cc}
                                                      M_1(p) & 0 \\
                                                      0& M_2(p)
                                                    \end{array}
  \Bigg),
\end{equation}  
with 
\begin{equation}
M_{1,2}(p) = \frac{M}{2} \left(1- \sqrt{1+ \frac{4 m^2}{M^2 F(p)}} \, \right).
\end{equation}

\begin{figure}[h]
\begin{minipage}{18pc}
\includegraphics[width=18pc]{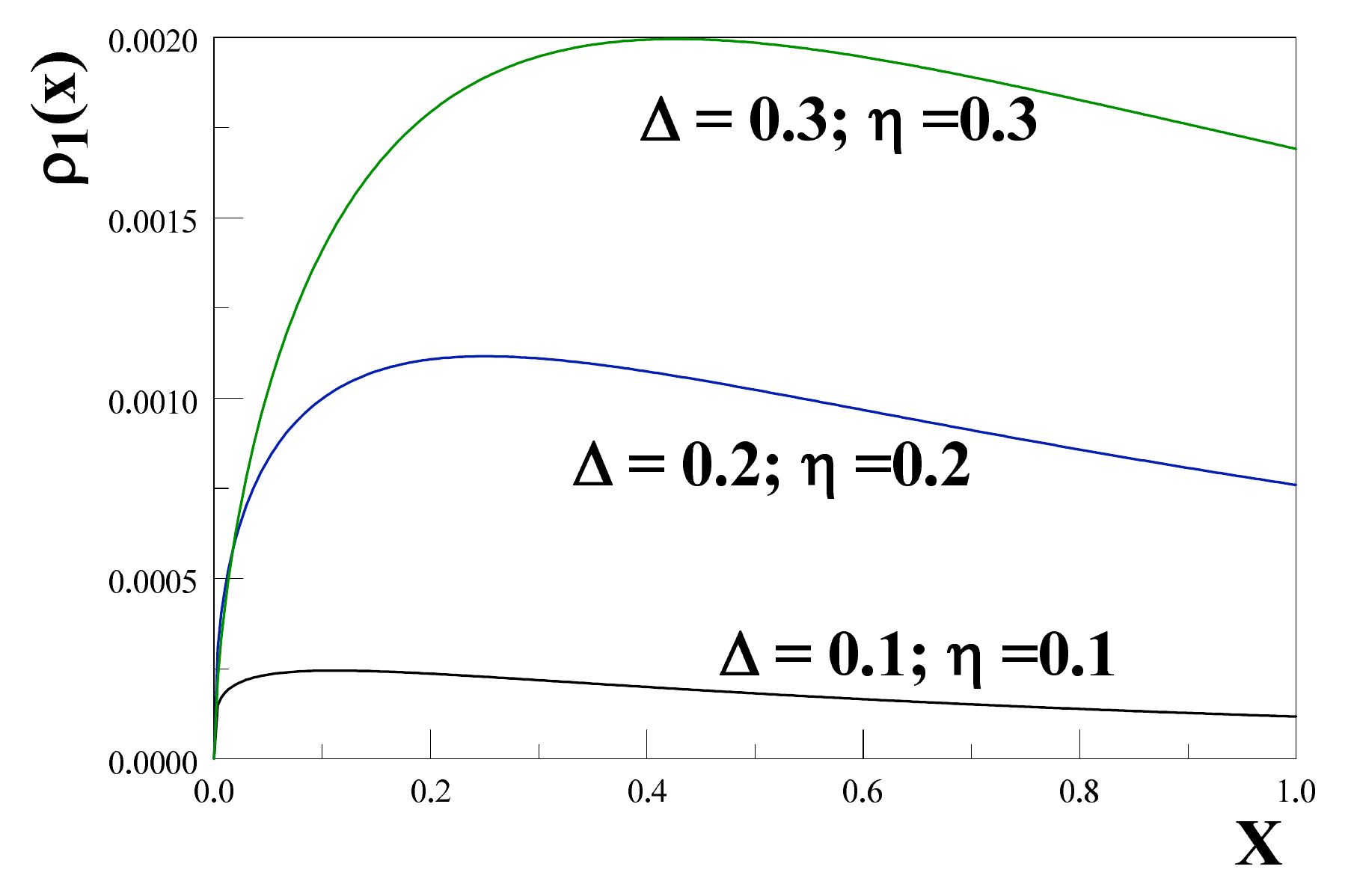}
\caption{Spectral density for the active-like mode.}
\label{fig:rho1}
\end{minipage}\hspace{2pc}%
\begin{minipage}{18pc}
\includegraphics[width=18pc]{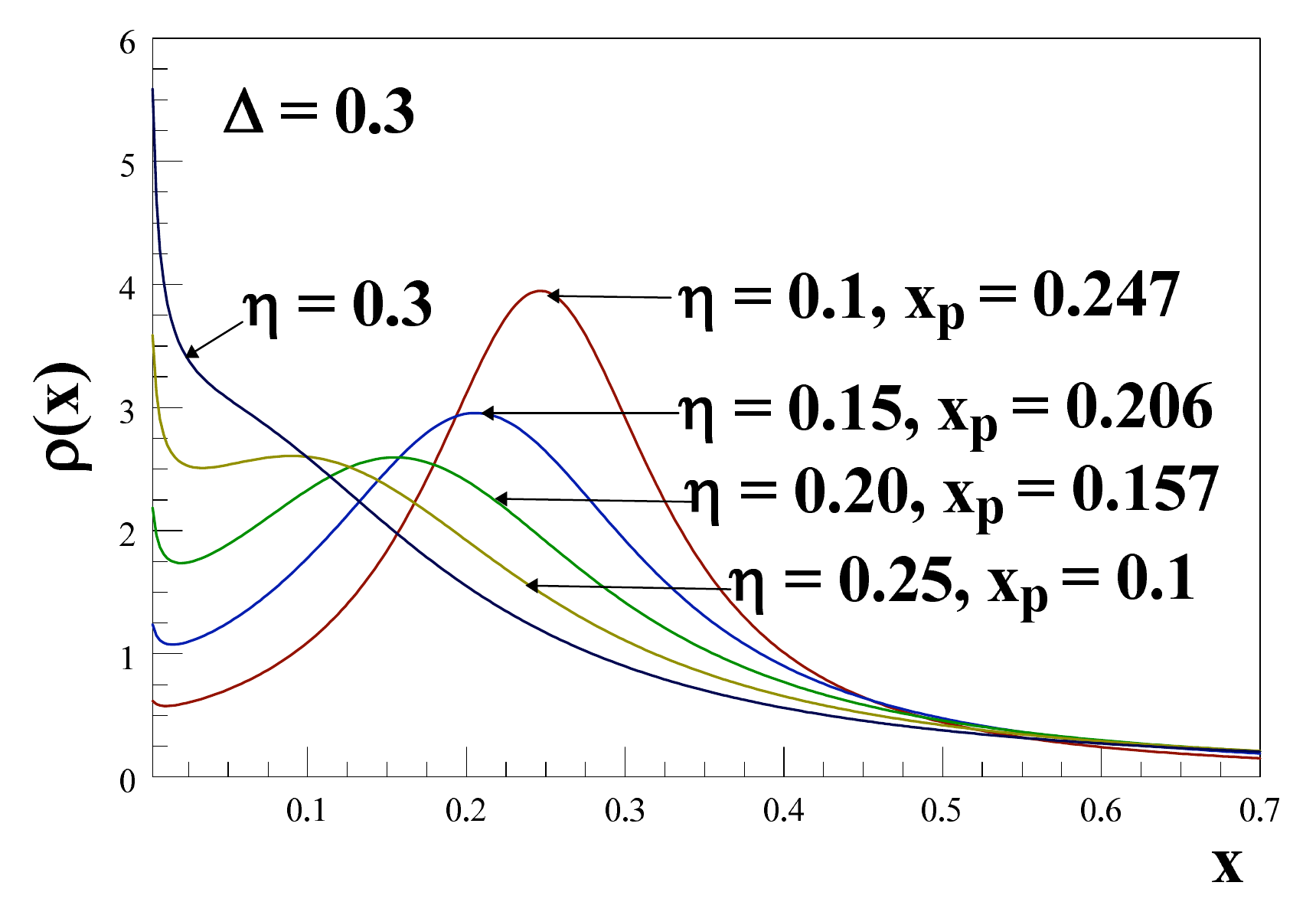}
\caption{Spectral density for the unsterile-like mode.}
\label{fig:rho2}
\end{minipage}
\end{figure}

The active-like ``mass" eigenstate has an isolated pole below the unparticle threshold $p^2=M^2$. This pole lies on the real axis and describes a massive and stable propagating mode. Near this pole, the propagator behaves like
\begin{equation}
\frac{1}{p^2 - M^2_1(p)} \approx \frac{Z_1}{p^2 - {\cal M}^2_1}, 
\qquad {\rm with} \qquad
{\cal M}^2_1 = \frac{m^4}{M^2} \Bigg[\frac{M^2}{\Lambda^2} \Bigg]^{2 \eta}
~~ {\rm and} ~~
Z^{-1}_1 \approx   1+ 2\eta ~ \frac{M^2_1}{M^2}.
\end{equation}
The active-like propagator also features an inherited spectral density
\begin{equation}
\rho_1(x) = \frac{\Theta(x)}{\pi} ~\frac{\frac{\Delta^2}{4}\,x^{2\eta}\sin(2\pi\eta)}
    {\Big[ x+1 - \frac{\Delta^2}{4}\,x^{2\eta}\cos(2\pi\eta) \Big]^2+\Big[\frac{\Delta^2}{4}\,x^{2\eta}\sin(2\pi\eta) \Big]^2},
\end{equation}
where
\begin{equation}
x = \frac{p^2-M^2}{M^2}
\qquad {\rm and} \qquad
\Delta = 2~\frac{m^2}{M^2} \Bigg[\frac{M^2}{\Lambda^2} \Bigg]^{\eta}.
\end{equation}
This spectral density is depicted in Fig. \ref{fig:rho1}. It vanishes at threshold $p^2=M^2$, increases rapidly reaching a broad maximum and diminishes for increasing $x$.

The dispersion relation for the unsterile-like ``mass" eigenstate has a solution only when $\mathrm{Re}(x) > 0$ and $0 \leq \eta < 1/3$. It describes a pole in the complex plane. Near this pole, the propagator behaves like
\begin{equation}
\frac{1}{p^2-M^2_2(p)} \approx \frac{Z_2}{p^2-{\cal M}^2_2+i{\cal M}_2\Gamma}, \label{prop2}
\end{equation}
where
\begin{equation}
{\cal M}^2_2 = M^2 \Bigg[1 + \Delta^\frac{1}{1-\eta}\, \cos\Big( \frac{\pi \eta}{1-\eta}\Big)\Bigg]
\qquad {\rm and} \qquad
\Gamma =\frac{ M^2}{{\cal M}_2}  \, \Delta^\frac{1}{1-\eta}\,\sin\Big( \frac{\pi \eta}{1-\eta}\Big). \label{M22}
\end{equation}

The spectral density for the unsterile-like mode is given by
\begin{equation}
\rho_2(x) = \frac{\Theta(x)}{\pi} ~\frac{ \Delta \,x^{\eta}\sin(\pi\eta)}{\Big[ x  - {\Delta }\,x^{\eta}\cos(\pi\eta) \Big]^2+\Big[{\Delta}\,x^{\eta}\sin(\pi\eta) \Big]^2},
\end{equation}
which is displayed in Fig. \ref{fig:rho2}. For $0 \leq \eta < 1/3$, there is a resonance with a maximum confirmed to be given by  
\begin{equation}
x_p = \Delta^\frac{1}{1-\eta}\, \cos\Big( \frac{\pi \eta}{1-\eta}\Big), \label{xpole}
\end{equation} 
as obtained in Eq. \ref{M22}.

In order to understand the significance of the imaginary part of the unsterile-like pole, let us revisit the RG resummation argument we used to derive the kinetic term of the unsterile field. To do that let us add
\begin{equation}
\mathcal{L}_{\rm m} = - \overline{\psi}_U ~ m ~\nu_a + \mathrm{h.c.}
\end{equation} 
to Lagrangian density \ref{hidden}. It is straightforward to diagonalize this action. The mass eigenstates are given by
\begin{eqnarray}
\psi_1 =  \nu_a \, \cos\theta_0 + \psi_U \, \sin\theta_0, \qquad &{\rm with}& \qquad M_1 \approx m^2/M, \nonumber \\
\psi_2  = - \nu_a\,\sin\theta_0 +  \psi_U\,\cos\theta_0,  \qquad &{\rm with}& \qquad M_2 \approx M + m^2/M,
\end{eqnarray}
where
\begin{equation}
\cos\theta_0 \approx 1 \qquad {\rm and} \qquad \sin \theta_0 \approx m/M.
\end{equation}
In terms of these mass eigenstates, the interaction between the unsterile and the ``hidden" conformal sector can be rewritten as
\begin{equation}
\mathcal{L}_{\rm int} = g\,\Big(\overline{\psi}_2\,\cos\theta_0-\overline{\psi}_1\,\sin\theta_0\Big)
 \mathcal{A} \Big({\psi}_2\,\cos\theta_0-{\psi}_1\sin\theta_0 \Big).
 \end{equation}

\begin{figure}[h]
\begin{center}
\includegraphics[width=12cm,keepaspectratio=true]{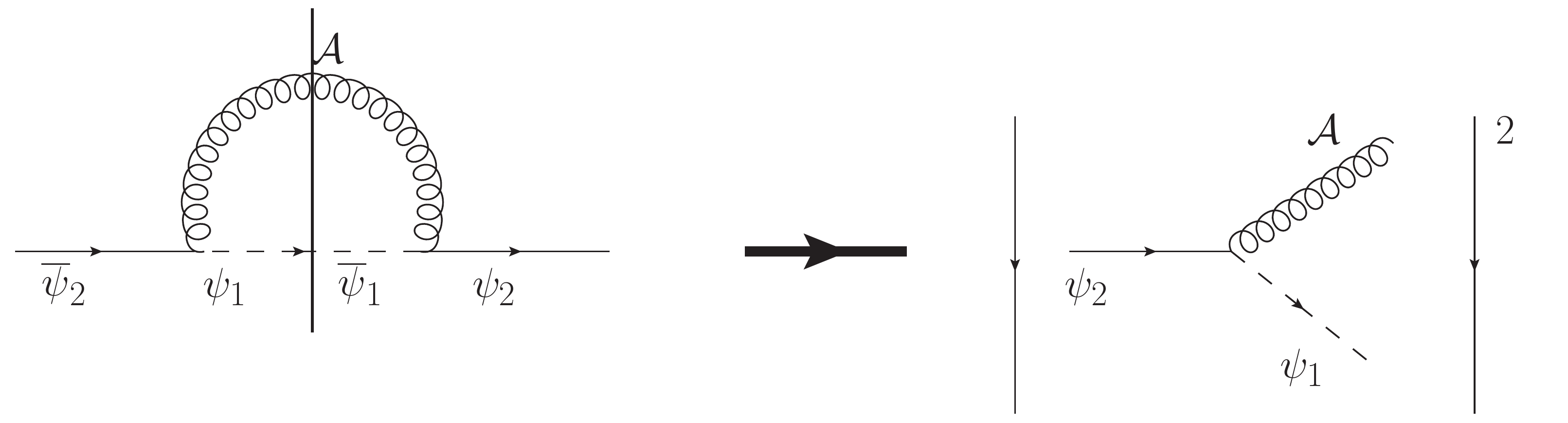}
\caption{Cutkosky cut for the self-energy of mass eigenstate $\psi_2$.}
\label{fig:se12cut}
\end{center}
\end{figure}

We can then obtain the decay rate for the process $\psi_2 \rightarrow \mathcal{A}\,\psi_1$ by applying Cutkosky cut to the self-energy diagram for the mass eigenstate $\psi_2$ as depicted in Fig. \ref{fig:se12cut}. It is given by
\begin{equation}
\Gamma_{\psi_2 \rightarrow \mathcal{A}\psi_1} = 2\pi \eta \,M_2\,\sin^2\theta_0 \, \cos^2\theta_0 \approx \pi \eta \frac{2m^2}{M}.
\end{equation}
To lowest order in $\eta$ and $m/M$, this result coincides with the non-perturbative imaginary part of the unsterile-like pole. Therefore, we see that this imaginary part $\Gamma$ describes the \emph{decay} of the unsterile-like mode into the active-like mode and fields in the ``hidden" conformal sector. 

\section{A Warm Dark Matter Candidate}
A motivation to consider sterile neutrino is that it is a potential warm dark matter candidate and could provide possible solutions to a host of astrophysical problems \cite{kusenko}. However, the radiative decay of a sterile-like neutrino mass eigenstate into an active-like mass eigenstate and a photon  leads to a decay line that could be observable in the X-ray or soft gamma ray background. Thus, the non-observation of this line provides a constraint on the mass and mixing angle of sterile-like neutrinos.

\begin{figure}[h]
\begin{center}
\includegraphics[width=12cm,keepaspectratio=true]{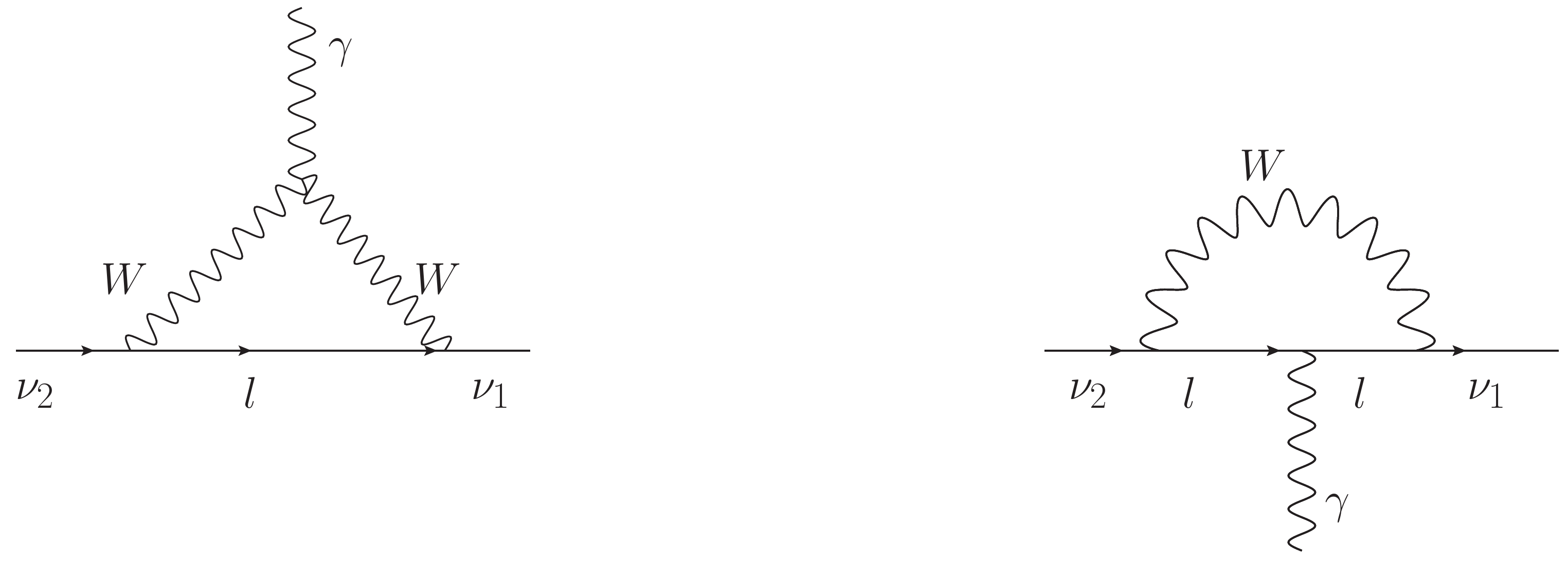}
\caption{Contributions to radiative decay of a sterile-like neutrino in unitary gauge.}
\label{fig:nu2tonu1gamma}
\end{center}
\end{figure}

Let us first consider the (canonical) sterile-active neutrino mixing
\begin{eqnarray}
\nu_a &=& \nu_1\,\cos\theta_0 + \nu_2\,\sin\theta_0, \nonumber \\
\nu_s &=&  - \nu_1\,\sin\theta_0 + \nu_2\,\cos\theta_0,
\end{eqnarray}
where $\nu_s$ is the canonical sterile neutrino. The charged current interaction then yields an interaction between the sterile-like neutrino and the charged lepton as follows
\begin{equation}
\mathcal{L}_{CC}= g\,\overline{\nu}_{aL}\, \not\!{W}\, l_L = g\, \Big( \overline{\nu}_{1L}\,\cos\theta_0 + \overline{\nu}_{2L}\,\sin\theta_0\Big)\, \not\!{W}\, l_L.
\end{equation}
This interaction vertex leads to the radiative decay of the sterile-like neutrino $\nu_2 \rightarrow \nu_1\,\gamma$ and the diagrams that describe this process in unitary gauge are
 shown in Fig. \ref{fig:nu2tonu1gamma}. The radiative decay width is given by
\begin{equation}
\Gamma_{\nu_2 \rightarrow \nu_1 \gamma} \approx \frac{\alpha_{em}}{2} \Big[\frac{3 G_F}{32\pi^2} \Big]^2 M^5_2~  \Big[\frac{m_l}{M_W}\Big]^4~\sin^2\theta_0\,\cos^2\theta_0.
\end{equation}

What about the radiative decay of the unsterile-like neutrino? In order to obtain the decay width for the unsterile case, we have to include the wave function renormalization $Z_{1,2}$. Furthermore, the mixing angles are accounted for by the following replacements
\begin{eqnarray}
\cos^2(\theta_0) &\rightarrow&  \frac{1}{2}\Big[1+  \widetilde{\mathcal{C}}(p)\Big]_{p^2=M^2_1},\nonumber \\
\sin^2(\theta_0) &\rightarrow& \frac{1}{2}\Big|1-  \widetilde{\mathcal{C}}(p)\Big|_{p^2=M^2_2}.
 \end{eqnarray}
The result is
\begin{equation}
\frac{\Gamma^U_{\nu_2 \rightarrow \nu_1 \gamma}}{\Gamma_{\nu_2 \rightarrow \nu_1 \gamma}}  =  \frac{\Delta^{\frac{\eta}{1-\eta}}~\Big[ \frac{M^2}{\Lambda^2}\Big]^\eta}{(1-\eta)\,(1+\eta \frac{\Delta^2}{2})}\approx \frac{ \Big[2\, \frac{m^2}{\Lambda^2}\Big]^\frac{\eta}{1-\eta}}{(1-\eta)\,(1+\eta \frac{\Delta^2}{2})}. \label{radratio2}
\end{equation}

We see that the unparticle nature of the sterile neutrino can lead to a \emph{substantial suppression} of the radiative decay rate. As an example, taking $m/M \sim 10^{-5}$, $M \sim \mathrm{keV}$ and $\Lambda \sim \mathrm{TeV}$, which are within the range of expectation for physics beyond the standard model, and taking $\eta \sim 0.1$, we find that the ratio \ref{radratio2} $\lesssim \mathcal{O}(10^{-3})$.
  
\section{Summary and Outlook}
In this talk, we considered the possibility that the $SU(2)$ singlet sterile neutrino might be an unparticle, an interpolating field whose correlation function feature an anomalous scaling dimension $\eta$ as a consequence of coupling to a ``hidden'' conformal sector. We studied the consequences of its mixing with an active neutrino via a see-saw mass matrix by focusing  on the simplest setting of one unsterile and one active Dirac neutrino. We found that there is no unitary transformation that diagonalizes the full propagator due to the non-canonical nature of the unsterile neutrino. This forces us to make a field redefinition for its complete diagonalization.

The active-like propagating mode corresponds to a stable particle, but inherits a non-trivial spectral density even in the absence of standard model interactions. The unsterile-like propagating mode is described by a complex pole above the unparticle threshold for  $0 \leq \eta < 1/3$, featuring an ``invisible width" that is a result from the decay of the unsterile-like mode into an active-like mode and fields in the conformal sector. 

We found that there is a substantial suppression of the radiative decay line width, resulting in a  weakening of the bounds from the X-ray and soft gamma ray backgrounds. This renders the unsterile neutrino as a good potential warm dark matter candidate. To further explore the possibility of an unsterile neutrino as a dark matter candidate, understanding its production process is necessary. Since an unsterile neutrino only interacts directly with the active one, the most effective dark matter production mechanism in this scenario is via unsterile-active neutrino oscillations. It would be interesting to study the implications of our results in the dark matter production mechanism along the line of Ref. \cite{dw}. 

We can also add another active neutrino into the mix, which results in the so-called $2+\tilde{1}$ model \cite{Boyanovsky:2009mq}, and study the dynamics of the active-active neutrino oscillations. The complications that come from the momentum-dependent mixing angles and non-trivial spectral densities render a full quantum field theoretical treatment of this model necessary. A work studying the active-active neutrino oscillations dynamics in the $2+\tilde{1}$ scheme using the real time quantum field theoretical formalism as introduced in \cite{Wu:2010yr}, in particular Section IV, is in progress. Perhaps the results will lead to the reconciliation of the LSND and MiniBooNE results.

\end{document}